\documentclass[12pt]{article}

\title{Charge density and electric charge in quantum electrodynamics}
\author{G. Morchio, \\ Dip. di Fisica, Universita' di Pisa and INFN, Pisa, Italy
\and F. Strocchi,
\\ Scuola Normale Superiore and INFN, Pisa, Italy}
\date{}

\makeatletter \@addtoreset{equation}{section}
\usepackage{latexsym}

\newtheorem{Definition}{Definition}[section]
\newtheorem{Proposition}{Proposition}[section]

\def \be {\begin{equation}}
\def \be {\begin{equation}}
\def \ee {\end{equation}}
\def \ume {{\scriptstyle{\frac{1}{2}}}}
\def \ra {\rightarrow}

\def \eqq {\equiv}

\def \a {{\alpha}}
\def \b {{\beta}}
\def \g {{\gamma}}
\def \d {{\delta}}
\def \eps {{\varepsilon}}

\def \l {{\lambda}}

\def \s {{\sigma}}

\def \t {{\tau}}

\def \o {{\omega}}

\def \A {{\cal A}}

\def \D {{\cal D}}
\def \F {{\cal F}}

\def \H {\mbox{${\cal H}$}}

\def \K {{\cal K}}

\def \O {{\cal O}}

\def \V {{\cal V}}

\def \Psio {{\Psi_0}}

\def \di {{\partial_i}}
\def \dj {{\partial_j}}
\def \dk {{\partial_k}}
\def \dl {{\partial_l}}
\def \do {{\partial_0}}
\def \dz {{\partial_z}}

\def \dmu {{\partial_\mu}}
\def \dnu {{\partial_\nu}}

\def \dr {{\partial_\rho}}
\def \ds {{\partial_\sigma}}

\def \do {{\partial_0}}

\def \jm  {j_\mu}

\def \abf {{\bf a}}

\def \k {{\bf k}}

\def \q {{\bf q}}

\def \x {{\bf x}}
\def \y {{\bf y}}

\def \z {{\bf z}}

\def \Rbf {{\bf R}}

\def \Nbf {{\bf N}}

\def \AO {{\cal A}({\cal O})}
\def \AO' {{\cal A}({\cal O}')}

\def \dxy {\delta(x-y)}
\def \at {{\alpha_t}}

\def \fR {{f_R}}

\def \] {\supseteq}

\def \Pf {{\bf Proof.\,}}

\def \limR {\lim_{R \ra \infty}}

\def \eqq {\equiv}
\def \Pf {{\bf Proof \,\,}}
\def \Rbf {{\bf R}}

\def\ra{\rightarrow}

\@addtoreset{equation}{section}

\def\AO'{\mbox{${\cal A}({\cal O}'$}}
\def\O{\mbox{${\cal O}$}}
\def \cal {\mathcal}
\def \o {{\omega}}
\def \O {{\mathcal O}}

\def \A {{\mathcal A}}
\def \AO {\A(\O)}

\def \F {{\cal F}}
\def \D {{\cal D}}
\def \H {{\cal H}}
\def \K {{\cal K}}

\def \psz {{\Psi_0}}
\def \psb {{\bar\psi}}
\def \a {\alpha}
\def \b {\beta}
\def \d {\delta}
\def \l {\lambda}
\def \t {\tau}
\def \g {{\gamma}}
\def \Cf {{C^\infty}}

\def \eps {{\varepsilon}}
\def \Dz {{{\cal D}_0}}

\def \dm {{\partial_\mu}}
\def \dn {{\partial_\nu}}

\def \di {{\partial_i}}
\def \dj {{\partial_j}}
\def \dk {{\partial_k}}

\def \be {\begin{equation}}
\def \ee {\end{equation}}

\def \yyyyy {\end{equation}}
\def \x {{\bf x}}
\def \y {{\bf y}}
\def \z {{\bf z}}
\def \k {{\bf k}}

\def \reali {\Rbf}

\def \o {{\omega}}
\def \O {{\Omega}}

\def \limR {\lim_{R \ra \infty}}
\def \QR {{Q_R}}
\def \QRT {$Q_R$}
\def \Dz  {$\D_0$}
\def \Du  {$\D_1$}
\def \dd {{\D \times \D}}
\def \DD {{\D \times \D}}
\def \DDz {{\D_0 \times \D_0}}
\def \DDu {{\D_1 \times \D_1}}
\def \QFP {{Q(\Phi, \Psi)}}

\def \ddt {$\D \times \D$}

\def \DDT {$\D \times \D$}
\def \DDzt {$\D_0 \times \D_0$}

\def \QRFP {(\Phi, \, \QR\, \Psi)}

\def \fp {{\Phi, \, \Psi \in \D}}
\def \FP1 {{\Phi, \, \Psi \in \D_1}}
\def \FP0 {{\Phi, \, \Psi \in \D_0}}

\begin{document}
\maketitle \pagenumbering{arabic} \makeatletter

\vspace{20mm}
\begin{abstract}
The convergence of integrals over charge densities is discussed in
relation with the problem of electric charge and (non local)
charged states in Quantum Electrodynamics (QED). Delicate, but
physically relevant, mathematical points like the domain
dependence of local charges as quadratic forms and  the time
smearing  needed for strong convergence of integrals of charge
densities are analyzed. The results are applied to QED and the
choice of time smearing is shown to be crucial for the removal of
the vacuum polarization effects responsible for the time
dependence of the charge (Swieca phenomenon). The possibility of
constructing physical charged states in the Feynman-Gupta-Bleuler
gauge as limits of local state vectors is discussed, compatibly
with the vanishing of the Gauss charge on local states. A
modification by a gauge term of the Dirac exponential factor which
yields the physical Coulomb fields from the Feynman-Gupta-Bleuler
fields is shown to remove the infrared divergence of scalar
products of  local and physical charged states, allowing for a
construction of physical charged fields with well defined
correlation functions with local fields.
\end{abstract}

\newpage
\section{Introduction}
The simple relation between charge density and electric charge in
classical electrodynamics does not extend trivially to the quantum
case, because of problems due to vacuum polarization and infinite
volume integration.

Quite generally, the relation between local charges and global
conserved charges has been extensively discussed in the seventies,
in relation with the proof of the Goldstone theorem ~\cite{O, R,
M, RQ} and it has become standard wisdom in the quantum field
theory (QFT) framework in which all the relevant information are
carried by the local states.

The problem changes substantially if the relevant charged states
are non local, as it is  the case of  Quantum Electrodynamics(QED)
~\cite{FPS}. As a consequence, one cannot rely on the standard
strategy of controlling the convergence of local charges on the
domain of local states, and in fact the limit of local charges, as
quadratic forms, crucially depends on the domain which is
considered. Moreover, as discussed by Swieca ~\cite{S}, on the
charged states obtained by applying Coulomb fields to the vacuum,
the local charge given by the integral of the density with the
standard smearing in space and time does not converge to the
electric charge and its limit is even time dependent.

This difficulty requires an analysis of the convergence of
suitably time-smeared integrals of the charge density; as we shall
see, not only the standard time smearing does not work, but also
Requardt's space-time smearing prescription ~\cite{RQ} requires a
modification in order to obtain the correct result for the
renormalized charge.  Actually, the basic point is the control of
the construction of charged states, which is related to the
infrared problem and is a deep non perturbative problem, both in
the general algebraic approach and in the approach which uses
fields operators ~\cite{B}.

Even in perturbation theory a rigorous control on the construction
of charged states is far from trivial. In the (positive) Coulomb
gauge the (non local) charged fields are difficult to handle
~\cite{SY} and the standard strategy is to use a local formulation
at the expense of positivity, as in the Feynman-Gupta-Bleuler
gauge. In this case, the charged states should be the obtained by
an appropriate construction in terms of local (unphysical) states.
Such a possibility has been advocated by Dirac and Symanzik
~\cite{D, SY} who proposed  explicit formulas for non local
charged (Coulomb) fields in terms of the local
Feynman-Gupta-Bleuler fields. Such a construction, which involves
non trivial ultraviolet and infrared problems has recently been
refined by Steinmann ~\cite{STE, SB} on the basis of a
perturbative expansion.

An important  issue is whether the above states can be constructed
only in terms of expectations of the observables or they exist as
vectors in a space in which local states are dense. In the latter
case, the control of limits of local states requires a topology
and the topology defined by the Wightman functions of the local
fields is too weak to give a unique space; thus, the possibility
of reaching the physical charged states, characterized by a
Coulomb delocalization, depends on the choice of a topology. For
example, the implicit use of the standard (Krein) metric on the
asymptotic fields $A_\mu^{as}$ excludes the presence of charged
states in the corresponding physical space, as pointed out by
Zwanziger in his investigations on the infrared problem in QED
~\cite{Z}. A possible non perturbative construction of  physical
charged states as limits of local states was discussed in Ref.
~\cite{MS}, with the use of a Hilbert-Krein topology which takes
into account the effects of the infrared problem. In our opinion,
the non uniqueness of such Hilbert-Krein majorant topologies,
which are associated to the Wightman functions of the local fields
in order to obtain weakly complete inner product spaces of states,
should not be regarded as a mathematical oddness, being related to
the allowed large distance behaviour or "boundary conditions" at
infinity.

The possibility of constructing  physical charged states as limits
of local state vectors in a weak topology has been recently denied
~\cite{SB} on the basis of an argument by which the local Gauss
charge, corresponding to the integral of $div\, E$, vanishes on
the local states and therefore on any weak closure of them; thus
no weak closure of the local states could contain physical charged
states. A main conclusion of our analysis is that the assumptions
involved in the argument underestimate the delicate r\^{o}le  of
such topologies for the convergence of local charges in QED.

In view of the problems which arise in QED, in Section 2 we
discuss in general charges defined  as limits of quadratic forms,
their crucial dependence on the domain and their relation to
global charge operators; in particular, attention is paid to the
case in which the relevant domains arise by applying non local
field operators to the vacuum.

In Section 3 we consider the problem of weak convergence of local
charges, which is shown to be very relevant for the Steinmann
argument. Strong convergence on the vacuum is shown to be a
general consequence of a stronger version of Requardt's theorem,
which also allows for an improved time smearing procedure,
necessary  for obtaining the correct value of the charge on
Coulomb charged states. Such a time smearing procedure avoids the
time dependence effects due to vacuum polarization while
preserving the correct value of the charge.

In Section 4, we discuss convergence of Gauss local charges on
physical charged states, on the basis of the standard local
formulations of QED, the Feynman-Gupta-Bleuler gauge. Quite
generally, independently of the use of a Hilbert-Krein topology,
it is shown that the construction of physical charged state
vectors as limits of local states in a weak topology is
incompatible with convergence, in the same weak topology, of the
Gauss local charges, even with a time smearing a la Requardt, on
local states. A simple model is discussed which mimics the
relation between charge density and charge in QED and displays the
compatibility between the vanishing of the Gauss charge on a dense
domain of local states and its strong convergence to a non zero
electric charge on the physical space. In Section 5 we compare the
construction of physical charged states of Ref.[12] with the DSS
construction analyzed by Steinmann ~\cite{STE, SB}. We show that
the infrared divergence  in the matrix elements of physical
charged states with local states is avoided by the modified DSS
exponential used in Ref. [12], which only differs from the
standard  factor by a gauge term. In this way one removes the
obstruction pointed out by Steinmann ~\cite{SB} as an argument for
the impossibility of constructing physical charged fields with
well defined correlation functions with local fields.


\section{Charges as limits of quadratic forms}

The analysis of the charge operator in QED presents subtle
features arising from the Coulomb delocalization of charged states
~\cite{FPS, MS}. It is therefore convenient to start by an
analysis of charges as integrals over a local density on general
(not necessarily local) domains.

In this section we shall show that i) charges defined as limits of
quadratic forms $Q_R$ on dense domains $\D \times \D$  in general
(including the quantum field theory case) crucially depend on the
domain; e.g. $Q_R$ may converge to zero on $\D \times \D$ and have
a non zero limit on $\D_1 \times \D_1,$ if $\D \cap \D_1$ is not
dense,  \, ii) such a phenomenon cannot occur if $Q_R$ converges
weakly on $\D$ and on $\D_1$.

Quite generally, in quantum field theory the problem of
associating an (unbroken) charge $Q$ to the integral over a local
density  $$Q_R = \int_{|x| \leq R}d x \,j_0(x, 0), \,\,\,\,\dm
j^\mu =0,$$ is delicate and deserves special attention.
Intuitively, one thinks of defining a  state of charge $q$, as
satisfying $$Q\,\Psi = \limR \QR \Psi = q \Psi, $$ but as
emphasized by Schoer and Stichel ~\cite{ST}, the limit does not
exist as a weak limit, even if some smearing in time is made with
$\a(x_0), \, \a \in \D(\Rbf)$ and even if $\Psi$ is a local state,
briefly $\Psi \in \D_0$. In the latter case, the limit exists
~\cite{M} as a sesquilinear form on \DDzt \, $$\limR \,(\Phi,
\,\QR \Psi) = Q(\Phi, \,\Psi), \,\,\,\Phi, \Psi \in \D_0.$$
Furthermore, if $\QR$ defines an unbroken symmetry on the local
fields the limit sesquilinear form defines an (hermitean) operator
$Q$ on $\D_0$.
\vspace{2mm}\newline i) {\bf Domains and limits of quadratic
forms} \vspace{1mm}\newline In general, the limit of  hermitean
operators $\QR$ as forms on domains $\D\times \D$, crucially
depends on the domain $\D$, in particular, the limit on $\D \times
\D$ does not constrain the limit on $ \D_1 \times \D_1$, $\D_1
\neq \D$.

Such a domain dependence in general persists, as shown by the
example below, even if $\QR$ converges to an hermitean
sesquilinear form $Q$ on $\dd$ satisfying the boundedness
condition \be{|\QFP | \leq C_\Psi \,||\Phi||, \,\,\forall \fp,}\ee
and therefore identifies an (hermitean) operator $Q$ with domain
$\D$. Furthermore, even if eq.(2,1) holds,  it is not at all
guaranteed that, $\forall\, \chi $, \be{(\chi, \,Q \Psi) = \limR
(\chi, \QR \,\Psi), \,\,\,\,\,\forall \Psi  \in \D.}\ee In fact,
such an equation means that $\QR \D$ converge weakly. By the
convergence of $\QR$ on $\dd$, weak convergence of $\QR\, \D$ is
equivalent to the boundedness of the norms $||\QR\,\Psi||$, for
each fixed $\Psi \in \D$.

In particular, as shown by the example below, even if $\QRFP$
converges to zero $\forall \fp $, one cannot conclude that
$\forall \chi$, $\limR (\chi, \QR \Psi)=0$. (This also shows that
the failure of eq.(2.2) does not depend on $\QR$ converging to an
unbounded or a bounded operator.)

The general phenomenon is that, if $\QR$ converge to an operator
$Q_0$ on $\D_0 \times \D_0$ and to an operator $Q_1$ on $\D_1
\times \D_1$ , the two operators $Q_0$ and $Q_1$ are in general
not related, in the sense of the following

\begin{Definition} Two densely defined hermitean operators $Q_0,
Q_1$ are said to be {\bf related} if there is an hermitean
operator $Q$ of which $Q_0$ and $Q_1$ are restrictions. They will
be said to be {\bf weakly related} if there is a densely defined
hermitean operator $Q_2$ to which both $Q_0$ and $Q_1$ are
related.
\end{Definition}
The above relations are symmetric and the second notion is
strictly weaker, since e.g. different self adjoint extensions
 of an hermitean operator are not restrictions
of the same hermitean operator. An example of limits of quadratic
forms which define not weakly related  operators is given below.

\vspace{1mm}{\bf Example}. Let us consider $L^2([0, \pi], d x ),
\,\,\D_0 \eqq$ the space of $ C^\infty$ functions vanishing at the
origin, $\D_1 \eqq$ the linear span of $f_1(x) \eqq 1$ and $f_n(x)
\eqq sin \,n x - \a_n sin\, x, \, n \geq 2 , \, \a_n \eqq 1/2
\int_0^{ \pi} sin \,n x$, so that $(f_1, \,f_n) =0$.

Clearly, both $\D_0$ and $\D_1$ are dense domains; in fact, if $f$
is orthogonal to $\D_1$ one has $$c_n \eqq (f, \,sin \, n x) =
\a_n (f, \, sin \,x) = \a_n c_1, \, n \geq 2.$$  Furthermore $$ 0
=(\pi/2) \int_0^\pi d x\, f(x)= \sum_{n\geq 1} c_n \int_0^\pi d x
\,sin \,n x = 2(\sum_{n\geq 2} \a_n^2 c_1 + c_1)$$ implies $c_1 =
0$, i.e. $f =0$. Now, let $\QR$ be the multiplication operator by
a regular function $q_R(x)$ converging to $\d(x)$  as a
distribution; then $$(\D_0, \QR \D_0) \ra 0,\,\,\,\, \,\,(\D_1,
\,\QR \D_1) \ra (\D_1, \, P_1 \D_1) \neq 0,$$ with $P_1$ the
projection on $f_1$. Thus, the limits of the hermitean operators
$\QR$ define two bounded operators which are not even weakly
related.

\vspace{1mm} Convergence of $\QR$  on $\D_0 \times \D_0$ to an
operator $Q_0$ constrains convergence to an operator on any domain
$\DD$, such that $\D \cap \D_0$ is dense.
\begin{Proposition} Let the hermitean operators $\QR$
converge to an operator $Q_0$ on $\DDz$ and to an operator $Q_1$
on $\DDu$; \newline i) if $\D_1 \cap \D_0 $ is dense, then $Q_0$
and $Q_1$ are weakly related \newline ii) if $\D_1 \supset \D_0$,
then $Q_0$ and $Q_1$ are related \newline iii) in both cases, if
$Q_0$ is essentially selfadjoint on $\D_1 \cap \D_0$, then $Q_1$
is contained in the closure of $Q_0$ \newline iv) if $Q_0$ and
$Q_1$ are not related, then $\QR$ does not converge to an operator
$Q$ on $\DD, \,\D= \D_0 + \D_1$.
\end{Proposition}
\Pf The hermiticity of $\QR$ implies that both $Q_i, \,i = 0,\,1$
are densely defined hermitean operators and so is  their
restriction $Q$ to $\D_1 \cap \D_0$. In case ii) $Q_1$ extends
$Q_0$, in case i) $Q_1$ and $Q_0$ extend $Q$. If $\QR $ converge
to $Q$ on $\D \times \D$, both $Q_0 $ and $Q_1$ are restrictions
of $Q$, so that $Q_0$ and $Q_1$ are related.

\begin{Proposition} If both $\QR \D_0$ and $\QR \D_1$ converge
weakly, then the two limits define  hermitean operators $Q_0$ and
$Q_1$ which are related
\end{Proposition}
\Pf Hermiticity of the limit forms follows from that of $\QR$ and
the existence of weak limits implies that the limit forms define
operators $Q_0$ on $\D_0$ and $Q_1$ on $\D_1$. The weak limit of
$\QR$ exists also on $\D_0 + \D_1$ and by the same argument
defines an hermitean operator $Q$ which extends $Q_0$ and $Q_1$.

As a result, if $Q_0$ is essentially selfadjoint, $Q_1$ is
contained in its closure and in particular if $Q_0= 0$, also $Q_1
=0$, in other terms if $\QR \D_0 $ converges to zero weakly and
$\QR \D_1$ converges weakly, then $Q_1 =0$.
\vspace{2mm}\newline ii) {\bf Convergence of local charges in
quantum field theory }\vspace{1mm}
\newline A general situation which occurs in quantum field theory is
described in terms of translational invariant (field) algebras
$\A_0, \,\A_1$, a (unique translationally invariant) cyclic vector
$\Psio$, domains $$ \D_0 = \A_0 \Psio, \,\,\,\,\D_1 = \A_1
\,\Psio, $$ and  local hermitean charges \QRT, with domains
containing  \Dz \,and \Du\, and with $(\Psio, \, Q_R\,\Psio) = 0$.
In general, \QRT\, is the integral of the zero component of a
local conserved (operator valued tempered distribution) current
$\jm$ with suitable smearing: \be{\QR = \int d^4x\, j_0(\x,
x_0)\,f_R(\x)\,\a(x_0) = j_0(f_R\,\a),}\ee $$f_R(x) = f(|x|/R) \in
\D(\Rbf^3), \,\,f(x) =1, \, if\, |x| \leq 1,\, f(x) = 0,\, if
\,|x| \geq 2,$$ $$ \a \in \D(\Rbf), \,\,\,supp \,\a \subset
[-a,\,a], \,a < 1 , \,\,\int d t \,\a(t) =1.$$

If $\A_0$ is a local (field) algebra and $(\D_0, \,\QR \D_0)$
converges as $R \ra \infty$, the limit defines an operator $Q_0$
iff \be{ \limR (\Psio, \,[ \QR, \, \A_0\,] \Psio) = 0, }\ee
equivalently ~\cite{M} iff  \be{ \limR (\D_0, \,\QR \Psio) =0.}\ee
Non local algebras may be relevant in the discussion of non local
states, e.g. asymptotic states, or  charged states in the Coulomb
gauge; a local and a non local field algebra, $\A_0$ and $\A_1$,
occur in the construction of charged states in QED.
\begin{Proposition} Let $\D = \A \Psio,$ $ \A $ an algebra
invariant under  translations; if on \ddt\,, $\QR$ converge to an
operator $Q$, then \be{\limR (\D, \QR \Psio) =
0.}\ee\end{Proposition} \Pf The spectral representations of the
space translations gives \be{((U(\abf) - 1 )^4 A \Psio, \,\QR
\,\Psio) = \int d\,J_A(\k) (e^{i\k  \cdot \abf} - 1)^4 R^3
\tilde{f}(R \k), \,\,\forall A \in \A}\ee where $\,\,d J_A(\k) =
\int d J_A(\k, k_0) \,\tilde{\a}(k_0)\,\,$ is a complex measure of
polynomial growth. Now, since for any polynomial $P(\k)$
$$|(e^{i\k \cdot \abf} - 1)^4\,R^3\, \tilde{f}(R \k) \,P(\k)|\leq
\frac{|R \k \cdot \abf|^4}{R} |\tilde{f}(R\k) P(R\k)|
|\frac{P(\k)}{P(R\k)}|\leq \frac{C}{R} \ra 0,$$ in the limit $R
\ra \infty$, the r.h.s. of eq.(2.6) converges to zero and
therefore, by the density of $\D$, one has   $$(U(\abf) - 1 )^4
\,Q \,\Psio = 0, \,\,\,\,\forall\, \abf.$$ Then, since $U(\abf)
-1$ is a normal operator, it follows that $(U(\abf) - 1 ) \,Q
\,\Psio = 0,$ and by the uniqueness of the translationally
invariant state $Q \Psio = \l \Psio $; actually $Q \,\Psio = 0$,
because $(\Psio, j_0 \Psio) =0$.

\vspace{2mm}Thus, under the same assumptions, one has that the
charge $Q'$ defined in terms of the limit of the commutator
~\cite{SWI}, coincides with $Q$, i.e. \be{ (\D,\, Q'\,A \Psio)
\eqq \limR (\D, \,[ \QR, \,A ]\, \Psio) = \limR \,(\D, \,\QR \,A
\Psio ).}\ee The domain dependence of charge operators obtained as
limits of quadratic forms appears also in the above quantum field
theory framework. In particular, as a result of Proposition 2.1,
if $\QR$ converges to zero on $\D_0 \times \D_0$, the convergence
to a non zero operator on $\D_1 \times \D_1 $ is excluded if $\D_1
\cap \D_0$ is dense, but may be allowed if $\D_1 \cap \D_0$ is not
dense, even if $ \Psio \in \D_1 \cap \D_0$.

Such features are illustrated and displayed by the following
Example.
\vspace{2mm}\newline {\bf Example.} Let $\phi$ be a massless
scalar field, $\psi$ a free Dirac field, $\A_0$ the algebra
generated by $\di \phi, \, i = 1, 2 , 3$ and by $\psi$ and $\A$
the algebra generated  by $\di \phi$  and  by $$ \psi_d(x) =
\psi(x) \,U(x), \,\,\,\,U(x) = e^{i\phi(f_x)}$$ $$ \phi(f_x)= \int
d y \,\phi(y) f(y - x ), \,\,\,f \in \D(\Rbf^4),\,\,\,\,\int d x
\,f(x) =1.$$ Then we consider the local charges $$ \QR^\phi \eqq
\,\partial_0 \phi(f_R\,\a,), \,\,\,\QR^\psi \eqq j_0 (f_R
\,\a),\,\, \,\,j_\mu (x) = :\bar{\psi} \g_\mu \psi:, $$ $$ \QR =
\QR^\psi + \QR^\phi$$ and the Fock representation  of $\psi,
\,\phi$, with Fock vacuum $\Psio$. Since by locality $$\limR
\,[\QR^\phi, \, \A_0] = 0, \,\,\,\limR (\D_0, \,\QR \,\Psio) =0,$$
we have  $$\limR (\D_0, \,\QR \,\D_0) = (\D_0,  \,Q^\psi
\,\D_0),$$ where $Q^\psi$ is the unbroken fermionic charge. On the
other hand, since $\lim_R [ \QR, \,\psi_d(g)] = 0$ we have $$
\limR \,(\D, \, \QR \,\D) = 0.$$ In conclusion $\QR$ converge to
the unbroken fermionic charge on $\D_0 \times \D_0$ and to the
zero charge on \DDT.

It is worthwhile to note that  the limit of the operators $\QR$
does not define an operator on $\D_{ext} \times \D_{ext}$, where
$\D_{ext} = \D_0 + \D$, (since the corresponding bilinear form is
discontinuous on the left). Moreover, one has a symmetry breaking
condition on the algebra $\A_{ext}$ generated by $\A_0 $ and
$\A_1$: $[\,\QR, \, \A_{ext}\,]\,\Psio$ converges weakly (actually
strongly) and $$\limR (\Psio, [\,\QR, \, \psi^\dagger \psi_d ]
\,\Psio) \neq 0. $$ This fact is actually a consequence of $Q_0$
and $Q_1$ being not related. In general if $\QR$ converges on
$\D_i \times \D_i,
 i = 0, 1$ to operators $Q_i$ which are not related, then, for the
algebra $\A$ generated by $\A_0$ and $\A_1$, one cannot have both
weak convergence of $[\,\QR, \, \A]\,\Psio$ and \be{ \limR (
\Psio, \,[\,\QR, \, \A]\,\Psio) = 0.}\ee  In fact, by
eq.(2.6),$$(\D_i,\, Q_i \,A \Psio) \eqq \limR \,(\D_i, \,[ \QR, \,
A\,] \Psio), \,\,\,\forall A \in \A_i.$$ Now, if eq.(2.9) holds,
by a standard argument ~\cite{SWI} one gets an hermitean operator
$Q$ on $\D \eqq \A \Psio$, which extends $Q_0$ and $Q_1$, in
contrast with their being not related.


\def \dzp {\D_0^{ph}}
\def \Hz {{\cal H}_0}
\def \dF {(\partial F)}
\def \dFR {(\partial F)_0(f_R \a_R)}
\section{Convergence of time smeared integral of charge density.
The vacuum sector of QED} In this section we discuss  weak and
strong convergence of local charges, in particular in the vacuum
sector of QED.

As found by Requardt ~\cite{R}, the weak limit of $\QR $ on local
states can be obtained under general conditions by a suitable time
smearing of the charge density, namely by considering, with $f_R,
\, \a$ as in eq.(2.3), \be{ \QR \eqq j_0(f_R\,\a_R), \,\,\,
\a_R(x_0) \eqq \a(|x_0|/R)/R.}\ee Actually, one can strengthen
Requardt's theorem and obtain strong convergence (Proposition
3.1), also with a more general time smearing $ \a_{T(R)}$, which
will prove necessary in the charged sectors of QED.

We recall that if $j_\mu$ is a  Lorentz covariant conserved
tempered  current, the two point function of the charge density is
of the form $$ < j_0(x)\,j_0(y) > = - \Delta\, J(x - y),$$ with
$J$ a Lorentz invariant tempered distribution of positive type; we
denote by $d \nu(k^2) $ the spectral measure defined by $J$.
\begin{Proposition}If the spectral measure $d\,\nu$ satisfies the
(infrared) regularity condition $$ d\,\nu(k^2) = k^2\,d\,\s(k^2),
\,\,\,\,d \,\s\,\,a \,\,\,measure,$$ then, putting $Q_{R, T(R)}
\eqq j_(f_R, \a_{T(R)})$ one has \newline i) $ s-\lim_{R \ra
\infty} \,Q_{R, R} \,\Psio = 0,$ \newline ii) $s-\lim_{R \ra
\infty} \,Q_{R, T(R)} \,\Psio = 0 $   for  all functions $T(R),$
with $T(R)/R \ra 0 $, satisfying $ T(R) > R^{1/3}$ and $R
\int_0^\eps d \sigma(s) \,s /( 1 + T(R)^2 s^2)^2 \ra 0 , \,\,\eps
> 0 $,
\newline iii) if, for $k^2 \in [0, \eps),\,\,\eps > 0$, $ d\,\s(k^2)/
d\,k^2 \leq C$, the above strong convergence to zero is obtained
by choosing $ T = R^{1/3 \,+\,\d}, \,\,\d >0$.
\end{Proposition}
\Pf In fact, one has $$||Q_{R,T}\,\Psio||^2 = \int d \nu(k^2)\,d^3
q \,\frac{| \q \tilde{f}(\q)|^2}{2 \sqrt{(|\q|/R)^2 + k^2})}\,R
\,|\tilde{\a}(T \sqrt{(|\q|/R)^2 + k^2})|^2. $$ Since $\a$ is of
fast decrease, $\forall \,N \in \Nbf,$ $$|\tilde{\a}(T
\sqrt{(|\q|/R)^2 + k^2})|^2 \leq \frac{C_N}{1 + ((T|\q|/R)^2 + T^2
k^2)^{N}} \leq \frac{C_N}{1+ (T^2 k^2)^N}, $$ and since $d \nu$ is
tempered there is an $M \in \Nbf$ such that $(1 + k^2)^{-M}\,d
\s(k^2) \eqq d \s'(k^2)$ is a finite measure. Then, by taking $N =
M + 2$, one has  $$ ||Q_{R,T}\,\Psio||^2 \leq C' \frac{R}{T} \int
d \s'(s^2)\, \frac{T s}{( 1 + T^2 s^2)^{2}} \eqq \frac{R}{T}
\,G(T).$$ The integrand function is bounded and converges to zero
pointwise, when $T \ra \infty$ , so that by the dominated
convergence theorem $G(T) \ra 0$. Thus i) is proved; moreover
strong convergence to zero holds if one chooses $R = T G(T)^{-1 +
\d}, \,
\d
>0$ and ii) follows since
$\forall \,\eps >0$, $$ \int_\eps^\infty d \s'(k^2) \,T
\sqrt{k^2}/(1 + T^2 k^2)^2 = O(1/T^3).$$ If the hypothesis of iii)
holds one can bound the integral from $0$ to $\eps$ by $$C
\int_0^\eps d s^2 \,\frac{T s}{(1 + T^2 s^2)^2} \leq \frac{C}{T^2}
\int_0^\infty d u^2 \frac{u}{ (1 + u^2)^2} = O(1/T^2).$$ Thus, the
strong convergence to zero is obtained if $T(R) = R^{1/3 + \d},
\,\d > 0$.

In the physical vacuum sector $\H_0$ of QED the assumptions of
Proposition 3.1 for the spectral measure of the electric current
are satisfied since $$ <
\partial F_0(x)\, \partial F_0(y) > = \int  k^2 d \rho(k^2)\, d^3
k\, |2 \sqrt{\k^2 + k^2}|^{-1}\, \k^2 e^{i k (x-y)}, \, \,\partial
F _\mu \eqq \partial^\nu F_{\mu \nu}, $$ with $d \rho(k^2)$ the
spectral measure of the two point function of $ F_{\mu \nu}$.
Hence, for $T(R)$ as in ii) of Proposition 3.1, \be{\limR||
\partial F_0(f_R, \a_{T(R)})\,\Psio||^2 =0}\ee and therefore $Q_{R, T(R)} =
j_0(f_R\,\a_{T(R)})$ converges strongly to zero on the dense
domain $\D_0^{ph}$ obtained by applying local bounded observable
operators to the vacuum. Eq.(3.2) with $T = R$ was also obtained
by D'Emilio ~\cite{DE}.

The situation is completely different if one adopts the standard
smearing ~\cite{O, R}, with a fixed $\a(x_0)$, \def \tQR
{\tilde{Q}_R} $$\tQR = j_0(f_R\,\a).$$

\begin{Proposition} The operators $\tQR$ have the following
properties \vspace{1mm} \newline i) they converge to zero on $\dzp
\times \dzp$ \vspace{1mm} \newline ii)   $ \tQR \,\Psio$ does not
converge weakly in $\H_0$, nor does $\tQR \, \Psi, \,\forall \Psi
= U \Psio$, $U$ a bounded local operator \vspace{1mm}
\newline iii) there are vectors $\Psi \in \Hz$ such that $$\limR
\, < \Psi, \tQR\,\Psio
>$$ depends on the time smearing test function $\a$ ({\bf time
dependence of the charge}) \vspace{1mm} \newline iv) there are
operators $F$ such that, $$\limR\, <\Psio,\,[ \tQR, \,F ] \Psio
> \neq 0 $$ (Swieca phenomenon ~\cite{S})
\end{Proposition}
\Pf  Since in the physical vacuum sector $\tQR = (\dF)(f_R \,\a)$,
i) follows by locality and Maison theorem ~\cite{M}.

For ii), the same calculation done above for $\QR$ now gives
$$||\tQR\,\Psio ||^2=  R \int  k^2\, d\rho(k^2)\,d^3 q \,\frac{|
\q \tilde{f}(\q)|^2}{2 \sqrt{(|\q|/R)^2 + k^2})}\,R \,|\tilde{\a}(
\sqrt{(|\q|/R)^2 + k^2})|^2, $$   so that $\tQR \Psio$ cannot
converge weakly. Furthermore, $\forall \Psi = U \Psio, $ $$ \tQR U
\Psio = [ \tQR, \, U ] \Psio + U \tQR \Psio $$ and the first term
on the r.h.s converges by locality; since the second term does not
converge weakly, neither does  the l.h.s.

In order to construct the vector $\Psi$ of iii) we consider
$$\Psi_R \eqq F_{0\, i}( (\di \Delta^{-1} g ) f_R \, h)\,\Psio,
\,\,\,\,g \in \D(\Rbf^3), \,\,h \in \D(\Rbf). $$ Such vectors
converge strongly to a vector $\Psi \in \Hz$, for $R \ra \infty$,
since the Fourier transform of $(\Delta^{-1}g)(\x)\,h(x_0)$ is
square integrable with respect to the measure  $d\,\rho(k^2) \,d^3
k \,|k_0|^{-1} |\k|^{2}\, k^2$ defined by the Fourier transform of
$< \dF_0 (x) \, \dF_0(y) >_0$. Then, we have $$\limR < \Psi,
\,\tilde{Q}_R \Psio > = \limR \,\int d \rho(k^2)\, d^3 k\,|2
k_0|^{-1}\, k^2 \tilde{f_R}(\k)\,\tilde{\a}(k_0)\,
\bar{\tilde{g}}(\k)\,\bar{\tilde{h}}(k_0) \ra $$ $$
\bar{\tilde{g}}(0)\, \int d\ \rho(m^2) \,m \,\tilde{\a}(m)
\,\bar{\tilde{h}}(m),$$ which displays  the dependence on $\a$.

The operators  $F_R \eqq F_{0\, i}( (\di \Delta^{-1} g ) f_R \, h
)\,$ converge strongly to an operator $F$ on the dense domain $
\A_L \,\Psio, \,\,\A_L = $ the algebra of strictly localized
(bounded) observables, since they converge strongly on $\Psio$ and
$ [\,F_R, \,A\,],\, A \in  \A_L, $ becomes independent of $R$, for
$R$ sufficiently large by locality. Then, we have $$ \limR <
\Psio, [ \tilde{Q}_R, \, F\,]\, \Psio > = \bar{\tilde{g}}(0)\,
\int d \rho(m^2) \,m \,(\tilde{\a}(m) \,\bar{\tilde{h}}(m) -
\tilde{\a}(-m) \,\bar{\tilde{h}}(-m))$$ which does not vanish in
general.

The vector $\Psi$ reflects the infrared behaviour of "dipole
states" of the form $\psi_c^\dagger(f)\,\psi_c(g) \,\Psio$, where
$\psi_c(g)$ is the electron field in the Coulomb gauge,
constructed, e.g., according to the Dirac-Symanzik-Steinmann
~\cite{SY, STE} prescription. Thus, in QED, even in the vacuum
sector, the naive idea of the charge as the integral of the charge
density gives rise to substantial problems because of vacuum
polarization effects which disappear only with a suitable time
smearing. The same problems arise in the charged sectors of the
Coulomb gauge, as  stressed by Swieca ~\cite{S}; they are a
general consequence of the non locality of the charged Coulomb
fields.

In general, the standard procedure, eq.(2.3), corresponds to
taking, in the corresponding correlation functions in momentum
space, the limit $\k \ra 0$ and gives a $\d$ function in $\o$ only
in expectations on local states. On the other hand, Requardt time
smearing corresponds to taking a limit $\k, \,\o \ra 0$ on the
light cone; in expectations on local states, it coincides with
that of the standard smearing and it is $\a$ independent. As
discussed in the Appendix, $\a$ independence does not hold on the
(non local) charged states of QED and therefore a modification of
Requardt's prescription is required for QED.


\section{Charge density and charge  in local formulations of
QED} The relation between charge density and charge presents
further subtle aspects in the charged sectors. As a consequence of
the local Gauss' law, charged states cannot be local. In this
section we discuss the limit of the Gauss charges $$Q_R^G =
(\partial F)_0 (f_R\,\a_R)$$ as quadratic forms on local and on
physical charged states in the Feynman-Gupta-Bleuler formulation
of QED and the implications on the possibility of constructing
physical state vectors as weak limits of local states.

In the Coulomb gauge, since the charged fields are not local, one
has to discuss the limit of local charges on domains obtained from
the vacuum by a non local algebra, giving rise to the problems
discussed in Sect.2.

Even in perturbation theory the control of the Coulomb gauge is
difficult and the standard strategy is to use a local formulation
at the expense of positivity; this is the case of the Feynman or
Gupta-Bleuler gauge. In this case, the charged fields and the
vector potential $A_\mu$ are local but their vacuum expectation
values cannot satisfy positivity; the corresponding Wightman
functions define an indefinite inner product space $\D_0 =  \F
\Psio,$ (with $\F$ the local field algebra), with inner product
denoted by $< . ,\, .  >$, which does not contains physical
charged states ~\cite{FPS, MS}.

As suggested by perturbation theory, non local physical charged
states may be obtained as suitable limits of  local unphysical
charged state vectors. A possible non perturbative construction of
physical charged state vectors along these lines was discussed in
~\cite{ MS}.

Quite generally, a crucial issue is that the definition and the
control of the limit of local charged state vectors requires a
topology; even in the positive case the weak topology on $\D_0$
defined by the seminorms $p_y (x) = |< x, \,y >|$, i.e. by the
Wightman functions is too weak; on the other hand,  the inner
product space $\D_0$ does not identify a unique Hilbert-Krein
majorant topology $\t$ ~\cite{MS} and one has different closures
$\K_\t = \overline{\D_0}^\t$. For the physical interpretation, the
relevant space is the physical subspace $\K'_\t \subset \K_\t$,
identified by a subsidiary condition (which in QED  selects gauge
invariant states) and different topologies may give rise to
isomorphic physical spaces.

In general, ~\cite{MS} the dependence of the space $\K'_\t$ on the
topology $\t$ should not be regarded as a mathematical oddness,
since different closures of $\D_0$ reflect different "boundary
conditions" at infinity. Even in the standard theory of unbounded
hermitean operators the local domain of $C^\infty$ functions of
compact support may allow different self adjoint extensions,
corresponding to different boundary conditions; in the physical
applications the choice of one instead of the other is dictated by
physical considerations ~\cite{MS, STR}.  In the QED case the lack
of non uniqueness reflects the physical fact that different
Hilbert-Krein topologies, defined by majorant inner products $( .
,\, . )$, correspond to different large distance behaviours of the
limit states, classified in particular by the velocity parameter
of their Lienard-Wiechert electromagnetic fields al large
distances ~\cite{MS}. Thus, the choice of the Hilbert-Krein
topology is governed by physical considerations since it
determines the class of vector states which one can constructively
associate to the Wightman functions, i.e. the corresponding
closure $\K$ of the vector space $\D_0$. For these reasons it
should not be a surprise that $\D_0$ may allow different
extensions.  Even in the algebraic approach the construction of
the charged states, which correspond to non local morphisms of the
algebra of observables, is not under sharp control and in any case
does not resolve the multiplicity associated to the large distance
behaviour ~\cite{Bu}.

\def \DP {\D^{ph}}
\def \HP {\H^{ph}}

The choice of the Hilbert-Krein topology in local formulations of
QED was discussed at length in ~\cite{MS} also in connection with
Zwanziger unsuccessful attempt to construct physical charged
states, as a result of a too restrictive Hilbert-Krein topology.

It has been argued ~\cite{SB} that the Gauss charge converges
weakly to zero on the local states as a consequence of the
vanishing of the Gauss charge commutators with local fields, and
that this prevents the construction of physical state with non
zero Gauss charge as limits  of local states. We shall examine the
weak points of this argument in order.

First, the vanishing of the Gauss charge commutators with local
fields implies  the vanishing of the Gauss charge as a
quadratic form on $\D_0 \times \D_0$, (see eq.(2.8) and the
Appendix). The vanishing of the Gauss charge on a closure of
$\D_0$ would follow (see Proposition 4.2 below) if one had weak
convergence of $Q_R^G \D_0$ in the topology which defines such a
closure of $\D_0$.

As we shall see the validity of such a property is not constrained
by the correlation functions of the local fields and does not hold
in general. Actually, (see the Example below and the following
Section) one may find a Hilbert-Krein topology $\t$ which avoids
the weak convergence of  $Q_R^G $  and allows for the construction
of physical charged state vectors.

The failure of the $\t$-weak convergence of  $Q_R^G \D_0$ should
not appear strange, since it involves a topology whose r\^{o}le is
merely that of linking  the physical non local charged states to
the unphysical local states. It should be stressed that the Gauss
charge $Q_R^G$ may well converge weakly or even strongly on a
dense domain $\D^{ph}$ of physical states, with respect to the
intrinsic Hilbert topology of the physical space. This means that
$\forall \Phi \in \H^{ph}, \,\Psi \in \D^{ph}$, (equivalently
$\forall \Phi \in \H', \,\Psi \in \D', $where $\H' $ denotes the
distinguished subspace of $\K$ satisfying the subsidiary condition
and $\D'$ a dense subspace of $\H'$), one has that  $$\limR \,<
\Phi, \QR \,\Psi > = \limR\, < \Phi, \QR^G \,\Psi >, \,\,\, \QR
\eqq j_0(f_R\,\a_R) $$ exists, equivalently \be{< \QR^G \Psi, \,
\QR^G \Psi
> = || \QR^G \Psi ||^2}\ee are bounded. This,
however, does not mean that $\QR \D'$ or $\QR^G \D'$ converge
weakly with respect to the Hilbert-Krein  closure $\K$, since weak
convergence in $ \K$ amounts to the boundedness of $$ || \QR^G\,
\Psi ||^2_{H K} \eqq(\QR^G \Psi, \,\QR^G \Psi ), $$ where $( . , .
)$ is the majorant inner product which defines the Hilbert-Krein
topology and the corresponding closure $\K$ of the local states
$\D_0$.

Actually, independently of any Hilbert-Krein  majorant, there is
a conflict between the  construction of the physical charged
states in terms of the Wightman functions of the local field
algebra $\F$ and the weak convergence of $\QR^G$ in the
corresponding extension $\D$ of $\D_0$. This difficulty is an
intrinsic one, since it only involves the Wightman functions of
$\F$ and the existence of the physical charged states in an
extension $\D$ of $\D_0$ compatible with the inner product $< , >$
defined by the Wightman functions, namely such that the sequences
of elements of $\D_0$ which define the extension, have convergent
inner products $< ,
>$ ~\cite{MS2}. No reference is needed to a Hilbert-Krein majorant topology,
even if, clearly, any Hilbert-Krein majorant defines a weak
extension. To clarify this point we introduce the following
\begin{Definition} Given two vector spaces $D_0$ and $D_1$, with
inner products $< , >^{(0)}$ and $< , >^{(1)}$, we say that $D_1$
can be realized in a weak  extension of $D_0$ if there exists an
inner product  vector space $\V$ containing a weakly dense inner
product subspaces isomorphic to $D_0$ and a subspace isomorphic to
$D_1$.
\end{Definition}

If $D_0$ and $D_1$ are defined by the vacuum correlation functions
of two field algebras $\A_0, \,\,\A_1$, the property of $D_1$
being realized in a extension of $D_0$ is implied by the existence
of joint vacuum correlation functions of $\A_0$ and $\A_1$. In the
case of local formulations of QED, if the correlation functions of
the physical  field algebra $\F_1$, e.g. of the field algebra  of
the Coulomb gauge, can be constructed in terms of the correlation
functions of the local field algebra $\F$, one has an extended
field algebra $\F_{ext}$ generated by $\F$ and $\F_1$, and $\D_1 =
\F_1 \Psio$ is realized in an extension of $\D_0 = \F \Psio$.

\begin{Proposition} Let $D$ be a non degenerate vector space
with inner product $< , >$, $D_0$ a weakly dense subspace and $D_1
\subset D$; let $\QR$ be hermitean  charges and \be{\limR < D_0,
\, \QR \,D_0 > = 0,}\ee \be{ \limR < D_1, \, \QR \, D_1 > \neq
0.}\ee Then, $ \QR \, D$ cannot converge in the  weak topology
defined by $< ,
>$ .

In concrete, if  physical charged states $\Psi$ may be obtained as
limits of the local states of $\D_0$ in a Hilbert-Krein topology
$\t$, i.e. they belong to a  (Hilbert-Krein)  extension $\D$ of
$\D_0$ and \be{\limR < \D_0, \, \QR^G \,\Psi > = < \D_0, \,Q \Psi
> \neq 0,\,\,\,\limR< \D_0, \, \QR^G \,\Psio > = 0, }\ee
then $\QR^G \D_0$ cannot converge weakly with respect to $\t$.
\end{Proposition} \Pf \,\,Since $D_0$ is dense and
$D$ is non degenerate, eq.(4.2) and weak convergence imply that
$\QR D_0$ converges weakly to zero. Thus $$ < w-\limR \QR D, \,
D_0
> = \limR < D, \, \QR \,D_0 > = 0$$ and again by the density of
$D_0$, $\QR \,D$ converges weakly to zero,  which is incompatible
with eq.(4.3).

By eqs.(4.4) and locality $  < \D_0, \,\QR^G\, \D_0
> \ra 0$ and therefore, by the density of $\D_0$, weak convergence implies $  \QR^G
\D_0 \ra 0$ and $$ < \D_0, \,Q \Psi
> = \limR < \D_0, \, \QR^G \Psi
> = \limR < \QR^G \D_0, \, \Psi > = 0.$$

Thus, the construction of physical charged states in a
Hilbert-Krein extension of $\D_0$ is incompatible with weak
convergence of the Gauss charge $Q_R^G$ on $\D_0$.

The failure of  weak convergence of $\QR^G \,\Psio$ gives rise to
the same problems and features discussed in Sect. 2; in particular
the domain dependence of the limits of $\QR^G$ allows the
vanishing of such a limit on $\D_0 \times \D_0$ compatibly with
its being non zero on a domain containing non local states (as are
the physical charged states).

\vspace{1mm}
\def \do {\partial_0}
A Hilbert-Krein topology which allows the construction of physical
char\-ged states, avoiding  the weak convergence of $\QR^G
\,\D_0$, was discussed in ~\cite{MS} in terms of the properties of
the asymptotic fields $\A_\mu^{as}$. The mechanism is clearly
displayed by the following
 \vspace{2mm} \newline {\bf Example}.
Let $\psi_0$ be a (canonical) free massive Dirac field and
$\phi_1, \, \phi_2$ two massless scalar fields satisfying the
following (equal times) commutation relations $$ [\,\phi_1, \,
\phi_2\,] = 0, \,\,\,[\,\pi_1, \,\pi_2\,] = 0 ,\,\,\,[\,\phi_i,
\,\pi_i\,] = 0, \,\,\,\,\pi_i \eqq \do \phi_i, \,\,i = 1,2, $$ $$
[\,\pi_1(\x), \phi_2(\y)\,] = [\, \pi_2(\x), \, \phi_1(\y)\,] = -
i \,\d(\x - \y).$$ Then, the fields $$\phi_\pm \eqq (\phi_1  \pm
\phi_2)/\sqrt{2},\,\,\,\,\,\pi_\pm \eqq (\pi_1 \pm
\pi_2)/\sqrt{2}, $$ \be{ \psi(x) \eqq U(x) \psi_0(x), \,\,\,U(x)
\eqq :e^{i\,\phi_2}:(x)}\ee satisfy
 the following commutators and anti commutators $$
[\,\phi_\pm(x), \,\phi_\pm(y) \,] = \pm i \,D(x - y),
\,\,\,[\,\phi_\pm(x), \,\phi_\mp(y) \,] = 0, $$ \be{[\,
\phi_{\pm}(x), \, \psi(y) \,] = \pm\, i D(x-y)
\psi(y),\,\,\,\,\{\,\psi(x), \,\bar{\psi}(y)\,\} = i S(x-y),}\ee
where $D, \,S$ are the standard commutator functions for massless
scalar and Dirac fields. Thus, $\phi_\pm$ and $\psi$  are local
fields.

Our field theory model is defined by the vacuum correlation
functions of the field algebra $\F$ generated by $\psi, \,\,\phi_1
$ and $\partial_\mu \phi_2, \,\mu = 0, 1, ...3$ and their Wick
products; such correlation functions do not satisfy positivity.

Now, we consider the following local charges \be{\QR^\phi \eqq \do
\phi_1(f_R \a_R), \,\,\,\QR \eqq j_0(f_R \a_R), \,\,\,\QR^G \eqq
\QR -\QR^\phi,}\ee where $$ j_\mu(x) = : \bar{\psi}\,\g_\mu
\,\psi:(x) = : \bar{\psi_0}\,\g_\mu \,\psi_0:(x).$$

The factorization of the correlation functions of $\psi_0$ and
$\phi_\pm$ implies that $Q_R$ converges to  an  unbroken (non
zero) "electron" charge in sense of quadratic forms on $\F\,
\Psio$  and in fact  the correlation functions with unequal
numbers of $\psi$ and $\bar{\psi}$ vanish. Actually, $Q_R \,\F
\Psio$ converges strongly with respect to any  Hilbert-Krein
topology chosen to turn $\F\,\Psio$ into a pre-Hilbert space,
provided it is a product over fermion and boson Fock spaces since,
by positivity of the correlation functions of $\psi_0$, \be{
\,||\QR \Psio||^2_{H K} = < \QR \Psio, \,\QR \,\Psio >  \ra 0.}\ee

The charge $\QR^G$ requires a quite different discussion. The
field algebra $\F$ is neutral under $\QR^G$ \be{ \limR [\,\QR^G,
\, \F \,] = 0.}\ee Therefore, putting $\D_0 \eqq \F\, \Psio$, by
the argument at the beginning of Sect.2, ii), one has $$ \limR  <
\D_0, \,\QR^G \,\D_0
> = \limR < \D_0, \,\QR^G \Psio
> =0.$$

In the  analogy with the local formulation of QED, the local
charge $\QR^G$ plays the r\^{o}le  of the Gauss charge, $\QR$
plays the r\^{o}le of the electron charge $j_0(f_R\,\a_R)$ and
$\QR^\phi$ plays the r\^{o}le of the longitudinal charge
$\partial_0 \partial A(f_R\,\a_R)$, all smeared in time a la
Requardt. As in the QED case the correlation functions of
$\QR^\phi$ vanish.

The relevant question is whether by taking suitable limits of the
local states of $\D_0$ one can construct the analog of the
physical charged states, i.e. states $\Psi$ satisfying the
following condition: \newline i) positivity, i.e.  $$< \Psi,
\,\Psi
> \,\,\geq 0.$$ ii) relativistic spectral condition \newline iii) vanishing
expectation of the "longitudinal" field $\partial_0 \phi_1$ $$ <
\Psi, \,\partial_0 \phi_1 \Psi > = 0,$$ iv) non zero Gauss charge,
i.e. $$\limR < \Psi, \, \QR^G \Psi > = \lim_R < \Psi, \, \QR \Psi
> \,\neq 0.$$ In the following, such states will be briefly
referred to as "physical" charged states.

Similarly to the QED case, the selection of states of $\D_0$
satisfying i) - iii) is obtained by means of a supplementary
condition \be{\partial_0 \phi_1^- \Psi = 0,}\ee which amounts to
the exclusion of $\phi_2$ components.

As in the QED case,  the subspace $\D_0' \subset \D_0$ satisfying
the subsidiary condition has zero electric charge; in fact one has
$\D_0' = \F_0' \Psio$, where $\F_0'$ is the field algebra
generated by $\phi_1$ and by the Wick products $$:\bar{\psi}
\Gamma\,\psi: = :\bar{\psi_0} \Gamma\,\psi_0:, $$ with $\Gamma$
any element of the algebra generated by the gamma matrices. The
problem is whether  physical charged states may lie in some
completion of $\D_0$; as one can easily guess the candidates for
the physical states are the free fermion states $\A_f \,\Psio,
\,\, \A_f =$ the algebra generated by $\psi_0$.

If one looks for a Hilbert-Krein completion $\K$ of $\D_0$ given
by a Krein  topology on the boson space, a sufficient  condition
for $\A_f \,\Psio$ belonging to $\K$ is that the Hilbert-Krein
majorant $( \,, \,)$ has a Fock structure and, at the level of the
two point function $< \phi_i \, \phi_j
>, \,\,i,\,j = 1, 2$, is
given by a measure  (in $k$ space) of the form $$\left(
\begin{array}{clcr} |\k|^2\,\b(|\k|) \,\,\,\,\,\,& 0
\\ 0 \,\,\,\,\,\,\,\,\, &  |\k|^{-2} \,\b(|\k|)^{-1}
\end{array} \right )\,\frac{d^3 k } {|\k|},\,\,\,\,$$ with $$ \b(|\k|)  \sim_{|\k| \ra
0} |\k|^{-2\d}, \,\, \b(|\k|) \sim_{|\k| \ra \infty} \,|\k|^{2\d},
\,\,\,\,\,\d\, > \,0.$$ This is in fact the condition which allows
the construction of the field $\phi_2$ and therefore of $U(x)$
from the derivatives $\partial_\mu \phi_2$, so that $\psi_0$ can
be recovered from $\psi$ . More generally, the metric leading to a
majorization may be chosen independently for each charged sector,
i.e. $\b$ may depend on the charge $q$.

It is instructive to discuss the relation between the existence of
charged states and the convergence properties of $Q_R^G$, which
play a crucial r\^{o}le in Steinmann argument. First $$||Q_R^G
\,\Psio||^2 \eqq  <  \QR^G\, \Psio, \, \QR^G\, \Psio > \ra  0,$$
i.e. $ s-\lim \, Q_R^G \,\Psio =0$ in the Hilbert topology defined
by the semidefinite Wightman two point function of $j_0 -
\partial_o\, \phi_1$, exactly as in the QED case (Sect. 3,
eq.(3.2)).

However, the weak convergence of $\QR^G \,\Psio$ in $\K$, i.e.
with respect to the Hilbert-Krein  space to which the physical
charged states belong, requires the boundedness of the norm $$||
\QR^G \,\Psio ||_{HK}^2 = ( \QR^G\, \Psio, \, \QR^G\, \Psio)= $$
$$\int d^3 k |\k|^{-1} |\k|^2\,\b(|\k|) \tilde{\a}(R k_0)|^2\,|R^3
\tilde{f}(R\k)|^2 = $$ $$\int  d^3 q \, |\q|\, \b(|\q|/R)
|\tilde{\a}(|\q|)|^2\, |\tilde{f}({\bf q})|^2   \sim R^{2\d}$$
which requires $\d \leq 0$. A similar calculation for the weak
convergence of $Q_R^G \psi\,\Psio$ in $\K$ involves the choice of
the majorization of the boson field correlations in the $q =1$
sector and requires $\d_{q=1} \leq 0$, whereas the existence of
physical states with charge $q = 1$  requires $\d_{q=1} > 0$.

In conclusion, in the space $\K$ defined by the above metric with
$\d \,> \,0$, there are two dense domains $\D_0 = \F \Psio$ and
$\D_1 = \F_1 \Psio$, with $\F_1$ the field algebra generated by
$\psi_0 $ and by $\phi_1, \,
\partial_\mu \phi_2$, with the properties: \newline 1) $\QR^G$
converges to the zero operator on $\D_0 \times \D_0$, \newline 2)
$\QR^G \Psio$ converges to zero strongly in the Wightman
(semidefinite) scalar product,  but it does not converge (even)
weakly in the extended space $\K$; moreover $Q_R^G $ on  local
charged states  does not converge weakly in $\K$
\newline 3) $\QR^G$ converges to the non zero "electron" charge on
$\D_1 \times \D_1$ \newline 4) $Q_R^G$ converges strongly on any
vector of $\D_1$ satisfying the supplementary condition
(eq.(4.11)), in the intrinsic Hilbert topology defined by the
Wightman functions.

\vspace{1mm}The  model  also displays the intrinsic conflict
between the construction of the physical charged states and the
weak convergence of $\QR^G$ in the extended space which contains
them; in fact, in the model divergences appear in the limit of
matrix elements $ < e^{i \phi_2} \Psio, \, Q_R^G \,\Psio >$. The
model also indicates that in QED the Gauss charge {\em converges
strongly} to the electric charge on a dense domain of physical
states (in the intrinsic Hilbert topology of the physical space),
a property which is not shared in general by local charges in
quantum field theories.


\section{Comments on the construction of physical charged states}

The construction of physical charged states in local formulations
of QED, like the Feynman-Gupta-Bleuler gauge, is a relevant issue
because it is strictly related to a non perturbative solution of
the infrared problem and provides theoretical support and
clarification of the standard perturbative calculations. In Ref.
[13], it is argued that physical charged states cannot be obtained
as weak limits of the local states, which are at the basis of the
perturbative  expansion, and that they can only be defined as
limits of morphisms of the algebra of observables. The arguments
for such a conclusion are on one side the convergence to zero of
the Gauss charge in any weak closure of the local states (the
weakness of such an argument was discussed in the previous
section) and on the other side the divergence of the matrix
elements between local states and the physical charged states
constructed according to the Dirac-Symanzik-Steinmann (DSS)
prescription. In this section we shall critically  examine the
latter  argument and show that a modification of the DSS
prescription along the lines discussed in Ref. [12], leads to
convergent results for the construction of physical charged state
vectors as weak limits of local states.

For this purpose, we adopt the general framework of Ref.[12] and
in particular we shall base the discussion on the following
assumptions: \vspace{1mm} \newline I) ({\em existence of
asymptotic limits of the vector potential}) the asymptotic limits
$A_\mu^{as}, \,\, as = in/out$, of $A_\mu$ exist as (covariant)
free fields with the local states in their domains \vspace{1mm}
\newline II) ({\em infrared coherence of "essentially local" states})
there are states $\Psi$, in a weak extension of $\D_0$, with $<
\Psi, \, \Psi
> \,\,> 0 $ having a decomposition into (improper) states
$\chi^\a$,with $< \chi^\a, \, \chi^\a > = 1$, which are coherent
states for $A_\mu^{in}$ (or for $A_\mu^{out}$)
\be{(A_\mu^{in})^-(k) \,\chi^\a = - \d(k^2)\,F_\mu^{\a, -}(k)
\,\chi^\a,}\ee $$k^\mu F_\mu^{\a, -}(k) = - e\, G(k), \,\,\,\,
G^(0) = 1,$$ with $G^(k)$ a real symmetric rotationally invariant
regular function.

For concreteness, the index $\a$, which labels the improper
states, can be thought as arising in the direct integral
decomposition  with respect to the spectrum of the electron
momentum $P_\mu^{ch}$. For non perturbative and perturbative
arguments, which support I) and II), we refer to Ref.[12].

We then introduce a function $F_\mu^\a(k)$, with $k F^\a(k) =  e\,
sign\,k_0\, G(k)$, determined by its restriction $F_\mu^{\a,
-}(k)$ to $C^- = \{\k, -|\k|\}$ and by the reality condition
$F_\mu^\a(k) = \overline{F_\mu^\a(-k)}$, and an operator valued
distribution $F_\mu(k)$, with $$[\,F_\mu(k), \, A^{in}\,] =
0,\,\,\,\, F_\mu^-(k) \,\chi^\a
 = F_\mu^{\a,-}(k)\,\chi^\a.$$ Then, the field \be{B_\mu^{in}(k) \eqq
A_\mu^{in}(k) - \d(k^2)\,F_\mu(k),}\ee defined on $\D_{\Psi} \eqq
\A^{in} \Psi$, $\A^{in}$ the field *-algebra generated by
$A_\mu^{in}$, satisfies \be{B_\mu^{in, -}(x) \,\Psi = 0.}\ee A
physical charged state $\Psi_{ph}$ is then obtained by putting
\be{ \Psi_{ph} = e^{ i e\, B_\mu^{in}(f^{\mu})} \,\Psi,}\ee
provided that the (real) function $f^\mu$ satisfies \be{\d(k^2)
k^\mu \tilde{f}_\mu(k) = i\, \d(k^2)\,G(k).}\ee This equation
corresponds to the Fourier transform of the Dirac condition
$\partial^\mu f_\mu(x) = \d^4(x)$ restricted to the light cone,
since $B^{in}$ is a free massless field, with ultraviolet
regularization provided by $G(k)$. Clearly, all solutions of the
Dirac condition $k^\mu \,\tilde{f}_\mu(k) = i G(k)$ are also
solutions of eq.(5.5).

Eq.(5.5) implies a singularity for $\tilde{f}_\mu$ of order at
least $1/k_0$ on the light cone and therefore the construction of
$\Psi_{ph}$, through eq.(5.4), involves the introduction of an
infrared cutoff in $f_\mu$. The point is whether its removal can
be done in the correlation functions of $e^{ i e\,
B_\mu^{in}(f^{\mu})} $ and local fields (i.e. Coulomb electron
fields  exist in the closure of the Gupta-Bleuler space) or only
in the expectation of observables on $\Psi_{ph}$.

Since by eq.(5.3) $\Psi$ provides a Fock representation of
$B^{as}$, the existence of $\Psi_{ph}$ in a Hilbert-Krein closure
of $\D_\Psi$ can be reduced to the finiteness of the two point
function $$< B_\mu^{in}(f_\mu) \Psi, \,
\eta\,B_\nu^{in}(f^\nu)\,\Psi > = ( B_\mu^{in}(f_\mu) \Psi, \,
B_\nu^{in}(f^\nu)\,\Psi ) = $$ \be{ \int d^4 k \,H^{\mu\,\nu}(k)
\,\d(k^2) \,\overline{\tilde{f}}_\mu(k)\,\tilde{f}_\nu(k) \eqq ||
\tilde{f}\,||^2_{H K},}\ee  where $\eta$ is the operator which
defines the corresponding Fock Hilbert-Krein majorant topology.
Such a majorization property implies that \be{
||\tilde{f}\,||^2_{H K} \geq \,\,|< B_\mu^{in}(f_\mu) \Psi, \,
B_\nu^{in}(f^\nu)\,\Psi
>|\, = }\ee $$= |\int d^4 k
\,g^{\mu\,\nu}\,\d(k^2)\,\overline{\tilde{f}}_\mu(k)
\,\tilde{f}_\nu(k)| = |< f, \, f >|$$ and therefore, in
particular, $f_\mu$ should be chosen so that  the indefinite
product $< f, \, f
>$ is finite.

The  DSS solution of $\,k^\mu \, \tilde{r}_\mu(k) = i\, G(k)$,
namely $\,\tilde{r}_i(k) = - i\, k_i G(k)\,|\k|^{-2},\,\,$
$\tilde{r}_0(k) = 0, $ does not work, since one obtains $$ < r, \,
r > = \int d^3 k \, |\k|^{-5} k_i\,k_j \,g^{i\,j}\,G(k)^2/2,$$
which is logaritmically divergent for $\k \ra 0$ and therefore, by
eq.(5.7), it excludes the convergence of $||\tilde{f}\,||_{H K}$
for any choice of a majorant Hilbert-Krein topo\-lo\-gy. This
corresponds to the divergence of the two point function $< \Psio,
\,\psi(x)\,\overline{\Psi}_p(y) \,\Psio
>$ pointed out by Steinmann (Ref.[13], Ch. 12, p.190).

However, as discussed in Ref. [12], a suitable choice of $f_\mu$
avoids the divergence of $<f, f>$ and allows for a finite
Hilbert-Krein norm.

In fact, all functions of the form $\tilde{f}_\mu(k) =
\tilde{r}_\mu(k) -i k_\mu \,\tilde{g}(\k)$ are solutions of
eq.(5.5). Since they differ from the DSS solution by a pure gauge,
they lead to the same expectations for all observables, but they
have different indefinite inner products: $$< f, \, f
> =  \int d^3 k \, G(\k)\,( G(\k) + 2 |\k|^2\,
  \,\tilde{g}) \,|\k|^{-3} /2$$ which
vanishes with the choice $ \tilde{g}(\k) = -  |\k|^{-2}  G(\k)/2$.
Such a choice gives \be{\tilde{f}_\mu = - i \overline{k}_\mu
\,|\k|^{-2}\,G(\k)/2, \,\,\,\,\,\overline{k} \eqq ( k_0, \, -
\k);}\ee the corresponding operator $A_\mu^{in}(f^\mu)$ describes
"zero norm" (unphysical) in-photons and their control depends on
the choice of the metric.

The above construction of charged states, based on eqs.(5.4)
(5.8), coincides with that of Ref.[12], apart from an infrared
convergent gauge term, since in eq.(91) of Ref.[12] for the
"infrared dressing" $U$ $$c_\mu(k) = (\sqrt{2}\,|\k|)^{-1}\,(a
k_\mu + b \bar{k}_\mu), \,\,\,\,a \,b = 1, $$ $$ \eta d^+(f) \eta
+ d^+(f) = (2|\k|)^{-1/2}\, a_\mu^+(k^\mu h + \ume\bar{k}^\mu
|\k|^{-2} G(\k)),$$ with $h(\k) = O(|\k|^{-2+\d}), \,\d \,> 0$.

It remains to characterize the conditions on the Hilbert-Krein
topology which give $||\tilde{f}\,||_{H K} < \infty$. For this
purpose, in the photon $k$-space we introduce four orthogonal four
vectors $\eps^1_\mu(k), \, \eps^2_{\mu}(k), \,k_\mu,
\overline{k}_\mu$, where $\eps^1_\mu(k), \, \eps^2_{\mu}(k)$ are
(transverse) polarization vectors. Thus, the most general rotation
covariant form of $H^{\mu\,\nu}(k)$ is $$H^{\mu\,\nu}(k)= \b(|\k|)
\,k^\mu k^\nu/2 |\k|^2 + \g(|\k|)\,\bar{k}^\mu \,\bar{k}^\nu/2
|\k|^2 + P^{\mu\,\nu}(k),$$ where $P^{\mu\,\nu}$ denotes the
projection on the transverse polarization. Then, since $$\sum_\nu
k^\nu k^\nu = 2 |\k|^2, \,\,\,\,\sum_\nu H^{\mu\,\nu}(k) \,k^\nu =
\b(|\k|)\,k^\mu,\,\,\, \sum_\nu H^{\mu\,\nu}(k) \,\bar{k}^\nu \,=
\g(|\k|)\bar{k}^\mu\,$$ positivity of the matrix $H^{\mu\,\nu}$
requires $\b, \,\g \,> \,0$. Furthermore, since the metric
$\eta(k)$ is given by $$(\eta^{-1}(k))^{\mu\,\nu} = \sum_\s
g^{\mu\,\s}\,H^{\s\,\nu}(k),$$ the condition $\eta^2 = 1$ requires
$\b\,\g = 1$. Thus, one gets $$||\tilde{f}\,||^2_{H K} = \int d^3
k\,G(\k)^2 (4 \,|\k|^3\,\b(|\k|))^{-1},$$ which is finite if
$\b(|\k|) \geq |\k|^{-\d},\,\,\,\,\d \,>\,0,$ for $\k \ra 0$. This
corresponds to the choice of the metric discussed in the Erice
lectures Ref. [12], especially pp. 323, 324, where one can also
find a characterization of the metric on the asymptotic fields
$A_\mu^{in}$ under general condition on the Fock structure of the
representation of $A^{in}$ given by $\Psi_{ph}$. Weak convergence
of the gauge term $\partial_0\,
\partial A(f_R\,\a_R)\,\Psi$, which is expected to govern the
weak convergence of $Q_R^G\,\Psi$, ($\Psi$ the "essentially local"
states at the basis of the construction), would require $\d \,\leq
0$, as in the Example of Section 4. In fact, one has
$$||\partial_0\,
\partial A(f_R\,\a_R)\,\Psi||_{H K}^2 = ||(B_R  + C_R)\,\Psi||_{H K}^2, \,\,\,\,B_R  \eqq
\partial_0\,\partial B(f_R\,\a_R),$$ $$ C_R \eqq - e/2 \int d^3 q \,G(\q/R)\,\tilde{f}(\q)\, [
\tilde{\a}(|\q|) + \tilde{\a}(-|\q|)].$$ Now, $||C_R \,\Psi||_{H
K} ^2$ remains bounded in $R$ and $$ ||B_R\,\Psi||_{H K}^2 = \int
d^4 k\,\theta(k_0)\,\d(k^2)\, k_0^2\, H^{\mu\, \nu}(k) k^\mu k^\nu
\,|\tilde{f}_R(\k)\,\tilde{\a}_R(k_0)|^2 =$$ $$ = \int d^3 q
\,|\q|^3\,|\tilde{f}(\q)\,\tilde{\a}(|\q|)|^2\, \b(|\q|/R)$$
diverges if $\d
> 0$.

\vspace{2mm} A similar discussion of the choice of the solution of
the Dirac condition, can be done for the DSS construction of the
physical fields in terms of the local Gupta-Bleuler fields . Again
the solution given by eq.(5.8) yields states which differ from the
DSS states by a gauge transformation $\exp{(i e \partial A (g))}$
and by the exponential $\exp{(i e [\,\partial A (g), \,A(r)\,])}$
of an infrared divergent phase, so that all the expectations of
observables coincide with those of the DSS solution. However, it
is easy to see that the above phase removes the divergence to
order $e^2$ of the scalar product $< \Psio, \, \psi(x)
\overline{\Psi}(y)\,\Psio
>$, pointed out by Steinmann (Ref. [13], p.190) as an evidence of
the claimed  impossibility of constructing physical charged state
vectors as weak limits of local states.


\appendix\section{Appendix}
\def \psz {{\Psi_0}}
\def \psb {{\bar\psi}}
\def \di {{\partial_i}}
\def \dj {{\partial_j}}
\def \diu {{\partial^i}}
\def \dk {{\partial_k}}
\def \dl {{\partial_l}}
\def \dm {{\partial_\mu}}
\def \dll {{\partial_\lambda}}
\def \dn {{\partial_\nu}}
\def \dr {{\partial_\rho}}
\def \ds {{\partial_\sigma}}
\def \dz {{\partial_0}}
\def \dzu {{\partial^0}}
\def \dzd {{\partial_0^2}}
\def \dmu {{\partial^\mu}}
\def \dnu {{\partial^\nu}}
\def \li {\triangle^{-1}}
\def \k {{\bf k}}
\def \x {{\bf x}}
\def \xp {{\bf x}^\prime}
\def \z {{\bf z}}
\def \amd {|{\bf a}|}
\def \amdd {|{\bf a}|^{-2}}
\def \amdq {|{\bf a}|^{-4}}
\def \y {{\bf y}}
\def \dA {{\partial_i A^i}}
\def \dAG {{\partial_i A^i_{GB}}}
\def \dmA {{\partial_\mu A^\mu}}
\def \dxy {{\delta(\x-\y)}}
\def \Oxq {{O(|\x|^{-4})}}
\def \aL {{\alpha_\Lambda}}
\def \aT {{\alpha_T}}
\def \fR {{f_R}}
\def \gLR {g_{\Lambda R}}
\def \gR {{g_R}}
\def \gi {{g_i}}
\def \gk {{g_k}}
\def \gm {{g_m}}
\def \hR {{h_R}}
\def \fRa {{f_R \, \alpha}}
\def \gRa {{g_R \, \alpha}}
\def \hRa {{h_R \, \alpha}}
\def \gia {{g_i \, \alpha}}
\def \giR {{g_{i R}}}
\def \hiR {{h_{i R}}}
\def \giRa {{g_{i R} \, \alpha}}
\def \hiRa {{h_{i R} \, \alpha}}
\def \hiRt {{\tilde h_{i R}}}
\def \hjRt {{\tilde h_{j R}}}
\def \Ft {{\tilde F}}
\def \fmn {{f_{\mu \nu}}}
\def \fij {{f_{ij}}}
\def \fiz {{f_{i 0}}}
\def \dmk {{d\mu(k^2)}}
\def \at {{\tilde \alpha}}
\def \xmu {{1\over |\x|}}
\def \zpamu {{1\over |\z+\a|}}
\def \xmyi {{1\over |\x-\y|}}
\def \xmymd {{1\over |\x-\y|^2}}
\def \xmzi {{1\over |\x-\z|}}
\def \xmd {{|\x|^{-2}}}
\def \xmq {{|\x|^{-4}}}
\def \g {{\gamma}}
\def \gmu {{\gamma^\mu}}
\def \gam {{\gamma_\mu}}
\def \gai {{\gamma_i}}
\def \gz {{\gamma_0}}
\def \Dm {{D_\mu}}
\def \Dmu {{D^\mu}}
\def \rR {{\rho_R}}
\def \er {{e_{ren}}}
\def \erp {{e\over 4 \pi}}
\def \Zti {{Z_3^{-1}}}
\def \Ld {{L^2}}
\def \Hu {{H^1}}
\def \OR {{\Omega_R}}
\def \LdOR {{L^2(\Omega_R)}}
\def \HLOR {{H^1 \oplus L^2 (\Omega_R)}}
\def \Cf {{C^\infty}}
\def \LR {\Lambda R}
\def \sv {{ \ \ , \ \ \  }}
\def \rlt {{\reali^3}}
\def \rlq {{\reali^4}}
\def \Ofa {{\O_{f_a}}}
\def \PPP {{< \Pdc \Pu \psz , }}
\def \ppz {{ \psz >}}
\def \ppt {{ \psz >_T}}
\def \DMU {(\Delta)^{-1}}
\def \ULR {U_{\Lambda R}}
\def \ji {j_i}
\def \Fiz {F_{i 0}}

\def \Psip {\Psi_{ph}}
In the standard case, locality and unitarity of space time
translations imply ~\cite{M} that, for expectation on local states
$\Psi$, eq.(2.8) applies and one has $$ \limR\,(\Psi, \,j_0(f_R,
\a)\,\Psi) = \limR\,(\Psi, \,j_0(f_R, \a_R)\,\Psi) = $$ \be{=
\lim_{T\ra \infty} \,\limR\,(\Psi, \,j_0(f_R, \a_T)\,\Psi).}\ee
Actually, the argument for the vanishing of $\,\limR \,< A \,
j_0(f_R, \a) > $ only uses locality and the property that the
Fourier transforms  of $< A \,j_0(x) >$, $ A $ local, are
measures. Perturbation theory indicates that this holds in the
Feynman-Gupta-Bleuler formulation of QED, where the vanishing of
$\limR \,< A \,\, \partial^i F_{i 0}(f_R, \a) >$ also follows from
the cluster property of the local fields in perturbation theory.
For charged states in QED, obtained through a DSS-like
construction, one may obtain sufficient localization properties so
that the matrix elements $< \Psip, \,j_0(\x,x_0)\,\Psip >$ differ
by the corresponding elements on local states, in the spacelike
complement of a double cone,  by corrections of order $|\x|^{-6}$,
uniformly in $|x_0| < T_0, \,T_0
> 0$ ~\cite{BDMRS}. However, the matrix elements of $j_i$ on such
states  decrease as $|\x|^{-2}$ ~\cite{BDMRS}: $$< \Psip,
\,j_i(\x, \,x_0)\,\Psip > = (e/4\pi) \int d^3 z \,\partial_i^{\x}
|\x - \z|^{-1}
\partial_0^2 K(\z,\,x_0) + O(|\x|^{- 4}) ,$$ where $K$ is the commutator function
of the electromagnetic field and eq. (A.1) does not hold for the
Gauss charge. In fact, one has $$ \lim_R < \Psip, \, \dz \diu F_{i
0} (\fR ,t) \Psip
>
 = \lim_R < \Psip, \, - \di j_i (\fR ,t) \Psip > = $$
$$ =  e \, \int d^3 z \,\dz^2   K(\z,t) = - \, e \, \int \o \, d
\o \, (\o  \tilde{K} )(0,\o) \, e^{i \o t}. $$ The vanishing of
the last expression for all $t$ would imply $$ \o \tilde{K} (0,\o)
= \lambda \, \delta(\o) $$ and therefore, by Lorentz covariance,
$$ \tilde{K}(k) = \lambda \, \eps(k_0) \, \delta(k^2), $$ i.e. a
free theory. Thus, the expectation value of the electric charge,
i.e. the electric flux at space infinity, in a charged state
defined by Coulomb charged fields is time dependent, even if its
time derivative vanishes at $t=0$ (by antisymmetry in $\o$). A
current $\ji$ with non--zero flux at infinity is therefore
present, \lq\lq induced\rq\rq\ by vacuum polarization effects.

The renormalized charge is given by the limit of the matrix
elements of the electric flux, with a suitable smearing  in time
($f_R, \,\a_T(R)$ as before) $$  \lim_{R \to \infty}
  < \Psip,\,\di F_{i 0} (\fR \, \a_{T(R)}\,\Psip >.    $$
In fact, by putting $$ \tilde{K}(k) = \int d\rho (m^2) \,
\eps(k_0) \, \delta(k^2 - m^2) \sv  $$ it follows $$  \lim_{R \to
\infty}
  < \Psip,\,\di F_{i 0} (\fR \a_{T(R)}\, \Psip > \  = $$
$$ = \, e \,
     \lim_{R \to \infty}
   \int d \rho(m^2) \, d^3k \,R^3 \, \tilde f(R\k) \,
   \tilde \alpha (T(R)\sqrt{\k^2 + m^2}) $$
$$= e \int  d\mu(m^2) \,d^3q\,\tilde{f}({\bf q})\,
   \tilde {\alpha} ((T(R)/R)\sqrt{\q^2 + R^2 m^2}).$$ Now,
for $m^2 > 0$, $(T(R)/R)\sqrt{\q^2 + R^2 m^2} > T(R) m$ and $(1 +
m^2)^M\, \tilde{\alpha}(T m)$ is bounded uniformly in $T$ by a
function of fast decrease and converges pointwise to zero. For
$m^2 = 0 $, the argument of $\tilde{\a}$ converges to zero if
$T(R)/R \ra 0$ and is equal to $|\q|$ if $T = R$ . Then, by the
dominated convergence theorem, if $T(R)/R \ra 0$ one gets $\l e$,
with $\lambda$ the $\rho$ measure of the point $m^2 = 0$, which is
one by the renormalization condition of the asymptotic
electromagnetic field. On the other hand, for $T(R) = R$ one gets
$$ e
   \lim_{R \to \infty}
   \int d^3q\, d\rho(m^2) \,\tilde{f}({\bf q})\,
   \tilde {\alpha} (\sqrt{{\bf q}^2 + R^2 m^2})\,
    $$ \be{= \lambda \, e \;
    \int d^3k\, \tilde \alpha (|\k|) \,
               \tilde f (\k)  \eqq \l\,e\, C(\a, f),}\ee
again by the Lebesgue dominated convergence theorem. Thus,
Requardt's prescription gives the renormalized charge up to a
factor $C(\a, f)$.

\end{document}